\documentclass[]{l4dc2023}

\usepackage{mathrsfs}
\usepackage{mathtools}

\usepackage{wrapfig}

\usepackage{pgfplots}
\usepgfplotslibrary{groupplots,dateplot}
\usetikzlibrary{patterns,shapes.arrows}
\pgfplotsset{compat=newest}
\usepackage{tikz}
\usetikzlibrary{shapes, arrows.meta, positioning}

\usepgfplotslibrary{groupplots}
\usepgfplotslibrary{fillbetween}
\usetikzlibrary{arrows,decorations.pathmorphing,positioning,fit,trees,shapes,shadows,automata,calc} 
\usetikzlibrary{3d,patterns,arrows,arrows.meta,calc,shapes,shadows,decorations.pathmorphing,decorations.pathreplacing,automata,shapes.multipart,positioning,shapes.geometric,fit,circuits,trees,shapes.gates.logic.US,fit, matrix,arrows.meta, quotes}
\usetikzlibrary{backgrounds,scopes}

\usepgfplotslibrary{external}
\pgfplotsset{compat=newest}
\newenvironment{customlegend}[1][]{%
	\begingroup
	\csname pgfplots@init@cleared@structures\endcsname
	\pgfplotsset{#1}%
}{%
	\csname pgfplots@createlegend\endcsname
	\endgroup
}%

\def\addlegendimage{\csname pgfplots@addlegendimage\endcsname}

\title[Physics-Informed Kernel Embeddings]{Physics-Informed Kernel Embeddings: \\ Integrating Prior System Knowledge with Data-Driven Control}

\usepackage{times}

\author{%
 \Name{Adam J. Thorpe}$^1$
 \thanks{These authors contributed equally to this work.}
 \Email{ajthor@unm.edu} \\
 \Name{Cyrus Neary}$^2$
 \footnotemark[1]
 \Email{cneary@utexas.edu} \\
 \Name{Franck Djeumou}$^2$
 \footnotemark[1]
 \Email{fdjeumou@utexas.edu} \\
 \Name{Meeko M. K. Oishi}$^1$ \Email{oishi@unm.edu} \\
 \Name{Ufuk Topcu}$^2$ \Email{utopcu@utexas.edu} \\
 \addr $^1$University of New Mexico\\
 \addr $^2$University of Texas at Austin%
}

\begin{document}

\maketitle

\definecolor{learnedWithSideInfo}{RGB}{0,128,0}
\definecolor{learnedWithoutSideInfo}{RGB}{4, 22, 184}
\definecolor{approxDynamics}{RGB}{150, 150, 150}
\definecolor{trueDynamics}{RGB}{0, 0, 0}

\newcommand{\canonicalCoord}{q}
\newcommand{\canonicalVelocity}{\dot{\canonicalCoord}}

\newcommand{\state}{x}
\newcommand{\stateSpace}{\mathcal{X}}
\newcommand{\control}{u}
\newcommand{\controlSpace}{\mathcal{U}}

\newcommand{\kernelFunction}{k}
\newcommand{\regularizationParameter}{\lambda}
\newcommand{\gaussianBandwidth}{\sigma}

\newcommand{\dynamicsFunction}{f}
\newcommand{\approxDynamicsFunction}{\Tilde{\dynamicsFunction}}

\newcommand{\numSamples}{M}
\newcommand{\sample}{\mathcal{S}}
\newcommand{\sampleOutput}{y}

\begin{abstract}

Data-driven control algorithms use observations of system dynamics to construct an implicit model for the purpose of control.
However, in practice, data-driven techniques often require excessive sample sizes, which may be infeasible in real-world scenarios where only limited observations of the system are available.
Furthermore, purely data-driven methods often neglect useful a priori knowledge, such as approximate models of the system dynamics.
We present a method to incorporate such prior knowledge into data-driven control algorithms using kernel embeddings, a nonparametric machine learning technique based in the theory of reproducing kernel Hilbert spaces. 
Our proposed approach incorporates prior knowledge of the system dynamics as a bias term in the kernel learning problem. 
We formulate the biased learning problem as 
a least-squares problem with a regularization term that is informed by the dynamics, that has an efficiently computable, closed-form solution.
Through numerical experiments, we empirically demonstrate the improved sample efficiency and out-of-sample generalization of our approach over a purely data-driven baseline.
We demonstrate an application of our method to control through a target tracking problem with nonholonomic dynamics, and on spring-mass-damper and F-16 aircraft state prediction tasks.

\end{abstract}
\section{Introduction}

The practical deployment of autonomous systems demands algorithms that can account for stochasticity and unexpected events due to humans in the loop or dramatic changes in the environment.
Model-based approaches to stochastic optimal control \citep{bertsekas1978stochastic, bertsekas2012dynamic} offer an analytic representation that is highly generalizable, but often rely upon strict model assumptions,
and can become inaccurate when deployed in new environments. 
They are particularly susceptible to model misspecifications, which can lead to inaccurate predictions that may lead to unpredictable or unsafe behaviors.
Data-driven control can account for poorly-characterized disturbances, but typically neglect prior system knowledge.
Additionally, these methods \citep{mauroy2020koopman, ansari2016sequential, rudy2017data} often exhibit poor data efficiency, meaning they require excessive sample sizes in order to adequately characterize the dynamical system behavior.

\begin{figure}
    \centering
    \vspace*{-0.5cm}
    \makeatletter
\tikzoption{canvas is xy plane at z}[]{%
    \def\tikz@plane@origin{\pgfpointxyz{0}{0}{#1}}%
    \def\tikz@plane@x{\pgfpointxyz{1}{0}{#1}}%
    \def\tikz@plane@y{\pgfpointxyz{0}{1}{#1}}%
    \tikz@canvas@is@plane}
\makeatother
\pgfplotsset{compat=1.15}
\pgfdeclareplotmark{fcirc}{%
          \begin{scope}[expand style={local frame}{\MyLocalFrame},local frame]
          \begin{scope}[canvas is xy plane at z=0,transform shape]
            \fill circle(0.12);
          \end{scope}   
          \end{scope}
}%
\tikzset{expand style/.code n args={2}{\tikzset{#1/.style/.expanded={#2}}}}
\newcommand{\GetLocalFrame}{
    \path let \p1=($(1,0,0)-(0,0,0)$), \p2=($(0,1,0)-(0,0,0)$),
    \p3=($(0,0,1)-(0,0,0)$) in \pgfextra{
    \pgfmathsetmacro{\ratio}{veclen(\x1,\y1)/veclen(\x2,\y2)}
    \xdef\MyLocalFrame{   
                x   =  {   (\x1,\y1)    },
                y   =  {    (\ratio*\x2,\ratio*\y2)     },
                z   =   {     (\x3,\y3)     }
            }
    }; }

\definecolor{learnedWithSideInfo}{RGB}{0,128,0}
\definecolor{learnedWithoutSideInfo}{RGB}{4, 22, 184}
\definecolor{approxDynamics}{RGB}{150, 150, 150}
\definecolor{trueDynamics}{RGB}{0, 0, 0}

\begin{tikzpicture}[
    scale=2,
    declare function={bivar(\ma,\sa,\mb,\sb)=
        1/(2*pi*\sa*\sb) * exp(-((x-\ma)^2/\sa^2 + (y-\mb)^2/\sb^2))/2;}]
    \begin{scope}[xshift=2.2cm, yshift=0cm]
        \draw[draw=learnedWithoutSideInfo!80, thick, rounded corners, fill=learnedWithoutSideInfo!0] (-0.5, -0.2) rectangle (1.35, 0.5);
        \begin{axis}[
            colormap/cool,
            at={(0.85cm, 0.25cm)},
            anchor={center},
            width=2.5cm,
            height=2.375cm,
            view={45}{65},
            enlargelimits=false,
            domain=-1:4,
            y domain=-1:4,
            samples=20,
            xtick=\empty, 
            ytick=\empty, 
            ztick=\empty,
            hide z axis,
            axis lines*=left,
        ]
            \addplot3 [surf] {bivar(2,1.75,2,1.25)};
        \end{axis}
        \node[align=center,  execute at begin node=\setlength{\baselineskip}{0.7em}] at (-0.05, 0.15) {\footnotesize \color{learnedWithoutSideInfo} Empirical\\ \footnotesize \color{learnedWithoutSideInfo}Distribution};
        
    \end{scope}
    \begin{scope}[xshift=2.2cm, yshift=0.75cm]
        \draw[draw=learnedWithSideInfo!80, thick,  rounded corners, fill=learnedWithSideInfo!0] (-0.5, -0.2) rectangle (1.35, 0.5);
        \begin{axis}[
            colormap/cool,
            at={(0cm, 0.25cm)},
            anchor={center},
            width=2.5cm,
            height=2.375cm,
            view={45}{65},
            enlargelimits=false,
            domain=-1:4,
            y domain=-1:4,
            samples=20,
            xtick=\empty, 
            ytick=\empty, 
            ztick=\empty,
            hide z axis,
            axis lines*=left,
        ]
            \addplot3 [surf] {bivar(2,1.75,2,1.25)};
        \end{axis}
        \node at (0.6, 0.15) {\Large +};
        \node[draw=approxDynamics!50!white, fill=approxDynamics!50!white, thick, rectangle, rounded corners] at (1, 0.15) {Bias};
        
    \end{scope}
    \begin{scope}[xshift=2.2cm, yshift=1.5cm]
        \draw[draw=approxDynamics!80!white, thick, rounded corners, fill=approxDynamics!0] (-0.5, -0.2) rectangle (1.35, 0.5);
        \node[draw=approxDynamics!50!white, fill=approxDynamics!50!white, thick, rectangle, rounded corners] at (0.9, 0.15) {Bias};
        \node[align=center,  execute at begin node=\setlength{\baselineskip}{0.7em}] at (0.1, 0.15) {\footnotesize \color{approxDynamics} Prior System\\ \footnotesize \color{approxDynamics}Knowledge};
    \end{scope}
    \begin{scope}[xshift=4.75cm,yshift=0.25cm]
        \begin{axis}[
            width=4cm,
            height=3.5cm,
            view={45}{65},
            enlargelimits=false,
            grid,
            minor tick num=1,
            xmin=-2,
            xmax=2,
            ymin=-2,
            ymax=2,
            zmin=0,
            zmax=1,
            clip=false,
            samples=10,
            xtick distance=0.25,
            ytick distance=0.25,
            xticklabels=\empty, 
            yticklabels=\empty,
            ztick=\empty,
            hide axis,
            axis lines*=left,
        ]
            \GetLocalFrame
            \begin{scope}[transform shape]
                \addplot3[only marks, draw=learnedWithoutSideInfo, fill=learnedWithoutSideInfo, mark=fcirc] coordinates {(0.84, -1.53, 0)};
                \addplot3[only marks, draw=trueDynamics, fill=trueDynamics, mark=fcirc] coordinates {(-0.25, 0.75, 0)};
                \addplot3[only marks, draw=learnedWithSideInfo, fill=learnedWithSideInfo, mark=fcirc] coordinates {(-0.15, 0.1, 0)};
                \addplot3[only marks, draw=approxDynamics, fill=approxDynamics, mark=fcirc] coordinates {(-1.15, -0.35, 0)};
                \node (R) at (axis cs:0.34, -0.73, 0) {};
                \node (B) at (axis cs:-0.25, 0.75, 0) {};
            \end{scope}
            \begin{scope}[canvas is xy plane at z=0]
                \draw [
                    thin, 
                    black!10,
                    step=0.25,
                ] (-2,-2) grid (2,2);
                \draw [
                    thin, 
                    black!20,
                    step=1,
                ] (-2,-2) grid (2,2);
            \end{scope}
        \end{axis}
    \end{scope}
    
    \draw[-Latex, thick, learnedWithoutSideInfo] (3.6, 0.125) ..controls (5.0,0.125) .. (5.68, 0.51);
    \node[align=left, execute at begin node=\setlength{\baselineskip}{0.7em}] at (4.15, -0.05) {\color{learnedWithoutSideInfo} \scriptsize Data-Driven\\ \color{learnedWithoutSideInfo} \scriptsize Kernel Embedding};
    
    \draw[-Latex, thick, learnedWithSideInfo] (3.6, 0.875) ..controls (5.25, 0.875) .. (5.85, 1.01);
    \node[align=left, execute at begin node=\setlength{\baselineskip}{0.7em}] at (4.15, 0.7) {\color{learnedWithSideInfo} \scriptsize Physics-Informed\\ \color{learnedWithSideInfo} \scriptsize Kernel Embedding};
    
    \draw[-Latex, thick, color=approxDynamics] (3.6, 1.625) ..controls (5.0, 1.625) .. (5.45, 1.15);
    \node[align=left, execute at begin node=\setlength{\baselineskip}{0.7em}] at (4.1, 1.8) {\color{approxDynamics} \scriptsize Prior Knowledge \\ \color{approxDynamics} \scriptsize Bias Term};
    
    \node[right, align=left] at (6.5, 0.3) {Reproducing Kernel \\ Hilbert Space (RKHS)};
    \draw[-Latex, thick] (6.75, 1.5) ..controls (6.5, 1.5) .. (6.15, 1.2);
    \node[right] at (6.8, 1.5) {\footnotesize True Embedding};

    
\end{tikzpicture}
    \caption{
        Physics-informed kernel embeddings combine data and prior system knowledge to more accurately estimate the expectation operator in an RKHS.  
    }
    \vspace*{-0.5cm}
    \label{fig:intro_figure}
\end{figure}
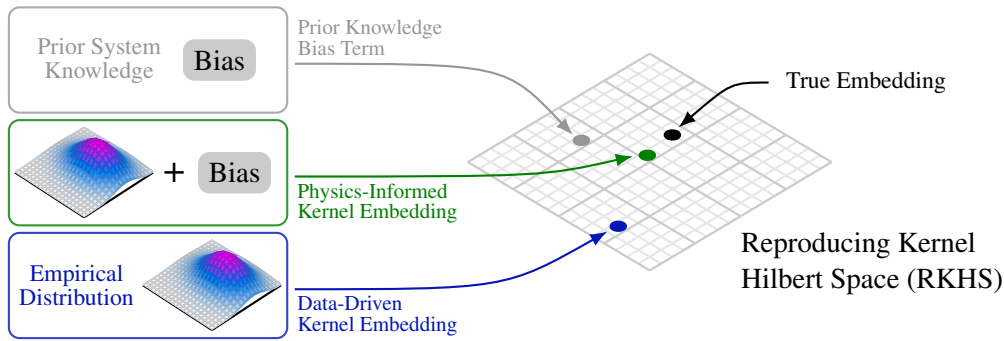

\textit{We present a method to incorporate (potentially) imperfect knowledge of the system dynamics in kernel embeddings in order to numerically estimate expectations in stochastic optimal control and state prediction problems.}
Specifically, we propose \emph{physics-informed kernel embeddings}, a nonparametric statistical learning technique based in reproducing kernel Hilbert spaces (RKHS) that incorporates prior knowledge of the dynamics as inductive bias.
As shown in \cite{thorpe2021stochastic, thorpe2022data, thorpe2022gradient}, data-driven reformulations of stochastic optimal control problems using kernel embeddings can efficiently be solved as a linear program by exploiting the mathematical properties of the RKHS.
However, despite the applicability to control, these techniques have thus far not seen widespread popularity, and presently do not take prior system knowledge into account.

We modify the regularized least-squares problem used to learn kernel embeddings with an additional bias term that encodes prior knowledge of the dynamics (Figure \ref{fig:intro_figure}). 
We present a representer theorem, which provides a closed-form solution to the learning problem.
Finally, we describe how the proposed physics-informed kernel embeddings may be applied to solve approximate stochastic optimal control problems.
We experimentally demonstrate our approach on state prediction and control tasks, including a spring-mass-damper system with a limited sample of system observations, a highly nonlinear F-16 aircraft, and a target tracking problem with nonholonomic dynamics.

\section{Related Work}
\label{section: related work}

Many approaches in data-driven control construct implicit black-box representations using
sparse regression over a library of nonlinear functions~\citep{kaiser2018sparse}, spectral properties of the collected data~\citep{proctor2016dynamic}, Koopman theory~\citep{abraham2017model, korda2018linear}, or Gaussian processes~\citep{krause2011contextual,pmlr-v120-gahlawat20a}. 
However, these approaches often suffer from high computational costs, expensive hyperparameter tuning, or nonconvexity of the surrogate functions, and are not readily amenable to incorporating prior dynamics. 

Methods to incorporate a priori knowledge into learned models of physical systems have been studied extensively over the past several years \citep[e.g.][]{pmlr-v168-djeumou22b,9930630, ahmadi2020learning}.
In particular, a number of recent works use neural networks to param\-etrize the unknown or unmodeled terms in differential equations \citep{djeumou2022neural, chen2018neural, rackauckas2020universal}.
This approach allows for the inclusion of general forms physics knowledge into data-driven models
, such as for so-called Lagrangian and Hamiltonian neural networks 
\citep{cranmer2020lagrangian, lutter2019deep, zhong2020unsupervised, allen2020lagnetvip, greydanus2019hamiltonian, matsubara2019deep, toth2019hamiltonian, finzi2020simplifying}, and it also enables learning control-oriented dynamics models \citep{zhong2019symplectic, zhong2020dissipative, roehrl2020modeling, duong2021hamiltonian, gupta2020structured, menda2019structured, zhong2021differentiable, shi2019neural}.
However, although these methods take advantage of physics-based knowledge, they often require extensive training data and training time.

Our proposed approach is based in the theory of kernel embeddings of distributions \citep{song2009hilbert, smola2007hilbert}, which have been applied to Markov models \citep{grunewalder2012modelling, nishiyama2012hilbert, song2010hilbert}, statistical inference \citep{song2009hilbert, song2010nonparametric}, policy synthesis \citep{pmlr-v38-lever15} and recently used to solve stochastic optimal control problems \citep{thorpe2021stochastic, thorpe2022data, thorpe2022gradient}. 
However, existing approaches to kernel-based control typically neglect prior knowledge of the system dynamics, or seek to encode structure directly into the kernel \citep{cheng2016learning} or learning prior \citep{geist2020learning}, which yields a highly specialized solution that does not generalize well to all systems or problem domains.
\section{Problem Formulation}
\label{sec:problem_formulation}


Let $(\mathcal{X}, \mathscr{B}_{\mathcal{X}})$ be a Borel space called the state space and $(\mathcal{U}, \mathscr{B}_{\mathcal{U}})$ be a compact Borel space called the control or input space.
We consider discrete-time stochastic systems of the form
\begin{equation}
    \label{eqn: system dynamics}
    x_{t+1} = f(x_{t}, u_{t}, \theta, w_{t}), \quad t = 0, 1, \ldots, N,
\end{equation}
where $x_{t} \in \mathcal{X}$ is the state of the system at time $t$, $u_{t} \in \mathcal{U}$ is the control action, $\theta \in \Theta$ are model parameters, and $w_{t}$ are independent random variables 
representing the stochastic disturbance.
The system evolves from an initial condition $x_{0} \in \mathcal{X}$ (which may be taken from an initial distribution $\mathbb{P}_{0}$ on $\mathcal{X}$).
For notational convenience, we can represent the dynamics in \eqref{eqn: system dynamics} via a stochastic kernel $Q : \mathscr{B}_{\mathcal{X}} \times \mathcal{X} \times \mathcal{U} \to [0, 1]$ that assigns a probability measure $Q(\cdot \mid x, u)$ to every $(x, u) \in \mathcal{X} \times \mathcal{U}$ on the measurable space $(\mathcal{X}, \mathscr{B}_{\mathcal{X}})$, as shown in \cite{bertsekas1978stochastic}. 

We presume the dynamics in \eqref{eqn: system dynamics} are unknown, meaning we do not have direct knowledge of the system dynamics or the uncertainty. Instead, we presume that a sample $\mathcal{S}$, consisting of observations taken independently and identically distributed (i.i.d.) from the system evolution is available, e.g. observations of the system transitions $\mathcal{S} = \lbrace (x_{1}, u_{1}, y_{1}), \ldots, (x_{M}, u_{M}, y_{M}) \rbrace$.
where $x_{i}$ and $u_{i}$ are taken from $\mathcal{X}$ and $\mathcal{U}$, respectively, and $y_{i} \sim Q(\cdot \mid x_{i}, u_{i})$. 
Such a sample may be collected from high-fidelity simulation or via observations of the system evolution from system trajectories. 
In addition, we presume that we have \emph{prior (potentially imperfect) knowledge of the dynamics}, $\tilde{f} : \mathcal{X} \times \mathcal{U} \to \mathcal{X}$.
Such prior knowledge may be available, for instance, if we only have access to a first-order  approximation of the dynamics, if the deterministic dynamics are available but the stochastic uncertainty is unknown, or if the model parameters $\theta$ are poorly estimated.

We solve, under the conditions above, two problems:
    1) 
    state prediction, where we seek to estimate the expected future state of the system after taking an action $u$ in a given state $x$, 
    \begin{equation}
        \label{eqn: prediction problem}
        \mathbb{E}_{y \sim Q(\cdot \mid x, u)}[y],
    \end{equation}
    and 2) 
    unconstrained stochastic optimal control, which can generally be written as
    \begin{equation}
        \label{eqn: stochastic optimal control problem}
        \min_{u \in \mathcal{U}} \quad \mathbb{E}_{y \sim Q(\cdot \mid x, u)}[c(y)], 
    \end{equation}
    where $c : \mathcal{X} \to \mathbb{R}$ is a (well-posed) arbitrary cost function that could capture, e.g.\ LQR, MPC, or other typical control objectives.
We focus on \eqref{eqn: prediction problem} and \eqref{eqn: stochastic optimal control problem} because they are representative of common problems in controls. 
As shown in \cite{thorpe2021stochastic}, by embedding the integral operator of the stochastic kernel $Q$ as an element in a high-dimensional space of functions known as a reproducing kernel Hilbert space (RKHS), we can approximate the expected value as a linear operation in the RKHS, and the approximate kernel-based reformulation of \eqref{eqn: stochastic optimal control problem} can be solved as a linear program. 
This is important because it provides a data-driven approach that is potentially amenable to run-time implementations. 
The main challenges are twofold: kernel embeddings neglect important information about the dynamics and are therefore more susceptible to errors, and they are susceptible to common sampling issues such as limited sample information.

\textit{The key contribution of this paper is a method to incorporate potentially imperfect knowledge of the system dynamics in the kernel embedding to numerically estimate \eqref{eqn: prediction problem} and \eqref{eqn: stochastic optimal control problem}.}
We propose \emph{physics-informed kernel embeddings}, that incorporates prior dynamics knowledge in kernel distribution embeddings, and apply our proposed technique to the problem of state prediction and control.
\section{Physics-Informed Kernel Embeddings of Distributions}
\label{section: biased kernel embeddings of distributions}


\subsection{Kernel Embeddings of Distributions}

Define the kernel $k : \mathcal{X} \times \mathcal{X} \to \mathbb{R}$, which is a positive definite function \cite[Definition~4.15]{steinwart2008support}. According to the Moore-Aronszajn theorem \citep{aronszajn1950theory}, given a positive definite kernel $k$, there exists a corresponding RKHS $\mathscr{H}$ of functions from $\mathcal{X}$ to $\mathbb{R}$
which satisfies the following properties:
    (i) 
    For all $x \in \mathcal{X}$, $k(x, \cdot) \in \mathscr{H}$, and 
    (ii)
    For all $f \in \mathscr{H}$ and $x \in \mathcal{X}$, $f(x) = \langle f, k(x, \cdot) \rangle_{\mathscr{H}}$, which is known as the reproducing property. 
Similarly, let $l : \mathcal{U} \times \mathcal{U} \to \mathbb{R}$ be a reproducing kernel over $\mathcal{U}$ and let $\mathscr{U}$ be its associated RKHS. 

Note that expectations $\mathbb{E}_{y \sim Q(\cdot \mid x, u)}[c(y)]$ are linear in the function argument $c$. 
As shown in \cite{grunewalder2012modelling}, assuming the kernel $k$ is measurable and bounded, there exists an element $m(x, u) \in \mathscr{H}$ called the \emph{kernel distribution embedding}, such that by the reproducing property, $\langle c, m(x, u) \rangle_{\mathscr{H}} = \mathbb{E}_{y \sim Q(\cdot \mid x, u)}[c(y)]$.
We can compute an empirical estimate $\hat{m}(x, u)$ of the embedding $m(x, u)$ using data $\mathcal{S}$.
As shown in \cite{grunewalder2012conditional}, the estimate $\hat{m}(x, u)$ can be computed as the solution to a regularized least-squares (RLS) problem,
\begin{equation}
    \label{eqn: regularized least squares problem}
    \hat{m} = \arg \min_{f \in \mathcal{V}} \frac{1}{2 \lambda} \sum_{i=1}^{M} \lVert k(y_{i}, \cdot) - f(x_{i}, u_{i}) \rVert_{\mathscr{H}}^{2} + \frac{1}{2} \lVert f \rVert_{\mathcal{V}}^{2},
\end{equation}
where $\mathcal{V}$ is a vector-valued RKHS of functions from $\mathcal{X} \times \mathcal{U}$ to $\mathscr{H}$ (see \citealp{grunewalder2012conditional} and \citealp{micchelli2005learning}) and $\lambda > 0$ is the regularization parameter. 
The solution to \eqref{eqn: regularized least squares problem} is given by a well-known class of theorems known as \emph{representer} theorems \cite{scholkopf2001generalized}. 


\subsection{Incorporating Prior Knowledge of the Dynamics in the Kernel Embedding}

Following \cite{scholkopf2001generalized}, we propose to learn a \emph{physics-informed kernel embedding} estimate via the following \emph{biased} RLS problem, 
\begin{equation}
    \label{eqn: biased regularized least squares problem}
    \hat{m}_{0} = \arg \min_{f \in \mathcal{V}} \frac{1}{2 \lambda} \sum_{i=1}^{M} \lVert k(y_{i}, \cdot) - f(x_{i}, u_{i}) \rVert_{\mathscr{H}}^{2} + \frac{1}{2} \lVert f \rVert_{\mathcal{V}}^{2} - \langle f, f_{0} \rangle_{\mathcal{V}},
\end{equation}
which differs from \eqref{eqn: regularized least squares problem} in that it includes an additional penalty term $\langle f, f_{0} \rangle_{\mathcal{V}}$, where $f_{0} \in \mathcal{V}$ is a user-specified bias term.
As discussed in \cite{scholkopf2001generalized}, this is a way to introduce bias into the regularization, and 
penalizes the difference between the learned function and the bias $f_{0}$, instead of only the RKHS norm $\lVert f \rVert_{\mathcal{V}}^{2}$.
The solution to \eqref{eqn: biased regularized least squares problem} can be characterized via a representer theorem, which we present as Theorem \ref{thm: biased representer theorem}.
\begin{theorem}
    \label{thm: biased representer theorem}
    If $\hat{f} \in \mathcal{V}$ minimizes the risk functional in \eqref{eqn: biased regularized least squares problem}, it is unique and has the form
    \begin{equation}
        \label{eqn: rls form}
        \hat{f} = \sum_{i=1}^{M} \beta_{i} k(x_{i}, \cdot) l(u_{i}, \cdot) + f_{0},
    \end{equation}
    where the coefficients $\beta_{i} \in \mathscr{H}$, $i = 1, \ldots, M$, are the unique solution of the set of linear equations,
    \begin{equation}
        \label{eqn: rls coefficients}
        \sum_{j=1}^{M} (k(x_{i}, x_{j}) l(u_{i}, u_{j}) + \lambda \delta_{ij}) \beta_{j} = k(y_{i}, \cdot) - f_{0}(x_{i}, u_{i}), \quad i = 1, \ldots, M.
    \end{equation}
\end{theorem}
\begin{proof}
    The proof is similar to \cite[Theorem~4.1]{micchelli2005learning}. 
    Let $f$ be any element of $\mathcal{V}$ such that $f(x_{i}, u_{i}) = k(y_{i}, \cdot)$, which minimizes the least-squared error of the data. 
    Let $g = f - \hat{f}$, and note that $\frac{1}{2} \lVert f \rVert_{\mathcal{V}}^{2}$ can be expanded as $\smash{\frac{1}{2} \lVert f \rVert_{\mathcal{V}}^{2} = \frac{1}{2} \lVert g + \hat{f} \rVert_{\mathcal{V}}^{2} = \frac{1}{2} \lVert g \rVert_{\mathcal{V}}^{2} + \langle g, \hat{f} \rangle_{\mathcal{V}} + \frac{1}{2} \lVert \hat{f} \rVert_{\mathcal{V}}^{2}}$.
    Let $\mathcal{E}(f)$ be the risk functional,
    \begin{equation}
        \label{eqn: empirical risk}
        \mathcal{E}(f) = \frac{1}{2 \lambda} \sum_{i} \lVert k(y_{i}, \cdot) - f(x_{i}, u_{i}) \rVert_{\mathscr{H}}^{2} + \frac{1}{2} \lVert f \rVert_{\mathcal{V}}^{2} - \langle f, f_{0} \rangle_{\mathcal{V}}.
    \end{equation}
    Taking the difference between the risk $\mathcal{E}(f)$ from \eqref{eqn: empirical risk} and the risk $\mathcal{E}(\hat{f})$ using \eqref{eqn: rls form} , and using the above expansion, we obtain
    \begin{align}
        \mathcal{E}(f) - \mathcal{E}(\hat{f}) ={}
            & \frac{1}{2 \lambda} \sum_{i} \lVert g(x_{i}, u_{i}) \rVert_{\mathscr{H}}^{2} - 2 \sum_{i} \langle k(y_{i}, \cdot) - \hat{f}(x_{i}, u_{i}), g(x_{i}, u_{i}) \rangle_{\mathscr{H}}
            \nonumber \\
            \label{eqn: proof step1}
            & 
            + \langle g, \hat{f} \rangle_{\mathcal{V}} + \frac{1}{2} \lVert g \rVert_{\mathcal{V}}^{2} - \langle g, f_{0} \rangle_{\mathcal{V}},
    \end{align}
    where the final term uses the fact that $\smash{\langle \hat{f}, f_{0} \rangle_{\mathcal{V}} - \langle f, f_{0} \rangle_{\mathcal{V}} = \langle \hat{f} - f, f_{0} \rangle_{\mathcal{V}} = - \langle g, f_{0} \rangle_{\mathcal{V}}}$. 
    Using the fact that for any $g \in \mathcal{V}$ and $f \in \mathscr{H}$, $\smash{\langle f, g(x, u) \rangle_{\mathscr{H}} = \langle g, k(x_{i}, \cdot) l(u_{i}, \cdot) f \rangle_{\mathcal{V}}}$, which comes from well-known properties of the vector-valued RKHS $\mathcal{V}$ \cite[Proposition~2.1]{micchelli2005learning}, and equations \eqref{eqn: rls form} and \eqref{eqn: rls coefficients}, we have that 
    \begin{align}
        \label{eqn: proof step4}
        \langle g, \hat{f} \rangle_{\mathcal{V}} - 2 \sum_{i} \langle k(y_{i}, \cdot) - \hat{f}(x_{i}, u_{i}), g(x_{i}, u_{i}) \rangle_{\mathscr{H}} = 0.
    \end{align}
    Then, using \eqref{eqn: proof step4} in \eqref{eqn: proof step1}, we have that
    \begin{align}
        \label{eqn: proof step5}
        \mathcal{E}(f) = \mathcal{E}(\hat{f}) + \frac{1}{2 \lambda} \sum_{i} \lVert g(x_{i}, u_{i}) \rVert_{\mathscr{H}}^{2} + \frac{1}{2} \lVert g \rVert_{\mathcal{V}}^{2} - \langle g, f_{0} \rangle_{\mathcal{V}} \geq \mathcal{E}(\hat{f}),
    \end{align}
    from which we conclude that $\hat{f}$ is the unique minimizer of $\mathcal{E}$ (uniqueness follows from the convexity of $\mathcal{V}$), which concludes the proof.
\end{proof}

In practical terms, Theorem \ref{thm: biased representer theorem} shows that the solution $\hat{m}_{0}$ to 
\eqref{eqn: biased regularized least squares problem} can be represented as a combination of two elements in the RKHS: a bias term $f_{0}$, and a linear combination of kernel functions $\sum_{i=1}^{M} \beta_{i} k(x_{i}, \cdot) l(u_{i}, \cdot)$ that represents the data-driven part.

Now it remains to choose a bias $f_{0}$.
A natural choice for $f_{0}$ is given by
\begin{equation}
    \label{eqn: bias term}
    f_{0}(x, u) = k(\tilde{f}(x, u), \cdot),
\end{equation}
such that for any $c \in \mathscr{H}$, $\langle c, f_{0}(x, u) \rangle_{\mathscr{H}} = \langle c, k(\tilde{f}(x, u), \cdot) \rangle_{\mathscr{H}} = c(\tilde{f}(x, u))$ by the reproducing property.
Then, using the solution $\hat{m}_{0}$ to the RLS problem in \eqref{eqn: biased regularized least squares problem} given by Theorem \ref{thm: biased representer theorem} and the bias term $f_{0}$ 
in \eqref{eqn: bias term}, 
we have that for any function $c \in \mathscr{H}$,
\begin{equation}
    \label{eqn: biased kernel distribution embedding}
    \mathbb{E}_{y \sim Q(\cdot \mid x, u)}[c(y)] \approx \langle c, \hat{m}_{0}(x, u) \rangle_{\mathscr{H}} = \boldsymbol{c}^{\top} W K(x, u) - \tilde{\boldsymbol{c}}^{\top} W K(x, u) + c(\tilde{f}(x, u)),
\end{equation}
where $\boldsymbol{c} \in \mathbb{R}^{M}$ and $\tilde{\boldsymbol{c}} \in \mathbb{R}^{M}$ are vectors with elements $\boldsymbol{c}_{i} = c(y_{i})$ and $\tilde{\boldsymbol{c}}_{i} = c(\tilde{f}(x_{i}, u_{i}))$, respectively, $\smash{W = (G + \lambda I)^{-1}}$, where $G \in \mathbb{R}^{M \times M}$, is a positive semi-definite matrix with elements $G_{ij} = k(x_{i}, x_{j}) l(u_{i}, u_{j})$, and $K(x, u) \in \mathbb{R}^{M}$ is a vector that depends on $x$ and $u$ that has elements $[K(x, u)]_{i} = k(x_{i}, x) l(u_{i}, u)$.

The estimate in \eqref{eqn: biased kernel distribution embedding} has a simple interpretation via addition and subtraction of the cost over the approximate dynamics from the expected cost, $\smash{\mathbb{E}_{y \sim Q(\cdot \mid x, u)}[c(y) - c(\tilde{f}(x, u))] + c(\tilde{f}(x, u))}$. Specifically, the first term $\boldsymbol{c}^{\top} W K(x, u)$ on the right-hand side of \eqref{eqn: biased kernel distribution embedding} 
corresponds to the purely data-driven kernel distribution embedding estimate; the second term $\tilde{\boldsymbol{c}}^{\top} W K(x, u)$ represents a kernel distribution embedding with the training data
$\lbrace (x_{i}, u_{i}, \tilde{f}(x_{i}, u_{i})) \rbrace_{i=1}^{M}$,
where we substitute the approximate dynamics over the data points $\tilde{f}(x_{i}, u_{i})$ for the observations $y_{i}$ in the dataset $\mathcal{S}$;
and the third term $g(\tilde{f}(x, u))$ is a correction that shifts the estimate such that it is centered around $\tilde{f}$.
\subsection{Control Using Physics-Informed Kernel Embeddings}
\label{section: application to kernel based control}

In this section, we demonstrate how physics-informed kernel embeddings can be used to solve the kernel-based control problem in \eqref{eqn: stochastic optimal control problem}. 
A stochastic policy $\pi : \mathscr{B}_{\mathcal{U}} \times \mathcal{X} \to [0, 1]$ for the system in \eqref{eqn: system dynamics} is a stochastic kernel that assigns a probability measure $\pi(\cdot \mid x)$ to every $x \in \mathcal{X}$ on $(\mathcal{U}, \mathscr{B}_{\mathcal{U}})$.
As shown in \cite{thorpe2021stochastic, thorpe2022data}, we can represent the stochastic policy $\pi$ as a kernel embedding $p(x)$ in the RKHS $\mathscr{U}$---a linear combination of kernels over a user-specified control set $\smash{\mathcal{A} = \lbrace \tilde{u}_{j} \rbrace_{j=1}^{P}}$, given by $\smash{p(x) = \sum_{j=1}^{P} \gamma_{j}(x) l(\tilde{u}_{j}, \cdot)}$, where $\gamma(x) \in \mathbb{R}^{P}$ are real coefficients that depend on the value of $x$. 

We use the physics-informed kernel embedding $\hat{m}_{0}$ (in place of the embedding $\hat{m}$) as in \eqref{eqn: biased kernel distribution embedding} to estimate the expected cost with respect to $Q$.  
Using $\hat{m}_{0}$ in \eqref{eqn: stochastic optimal control problem},
the policy embedding $p(x)$ can be found as the solution to the following problem,
\begin{subequations}
\label{eqn: approximate stochastic optimal control problem}
\begin{align}
    \min_{\gamma(x) \in \mathbb{R}^{P}} \quad & \boldsymbol{c}^{\top} W R(x) \gamma(x) - \tilde{\boldsymbol{c}}^{\top} W R(x) \gamma(x) 
    + C(x)^{\top} \gamma(x) \\
    \text{s.t.} \quad & \boldsymbol{1}^{\top} \gamma(x) = 1, \quad 0 \preceq \gamma(x)
\end{align}
\end{subequations}
where $\boldsymbol{c} \in \mathbb{R}^{M}$ 
and
$\tilde{\boldsymbol{c}} \in \mathbb{R}^{M}$ are as in \eqref{eqn: biased kernel distribution embedding}, $W \in \mathbb{R}^{M \times M}$ is a real matrix as in \eqref{eqn: biased kernel distribution embedding},
$R(x) \in \mathbb{R}^{M \times P}$ is a real matrix that depends on $x$, with elements $[R(x)]_{ij} = k(x_{i}, x) l(u_{i}, \tilde{u}_{j})$, and $C(x) \in \mathbb{R}^{P}$ is a vector with elements $[C(x)]_{j} = c(\tilde{f}(x, \tilde{u}_{j}))$. 
Notably, the problem in \eqref{eqn: approximate stochastic optimal control problem} is a linear program, and can be solved efficiently.
According to \cite{boyd2004convex}, since we seek to minimize a linear combination by choosing non-negative weights, it is immediately clear that we should allocate as much weight as possible to the smallest terms. Thus, the solution is a vector $\gamma^{*}(x) \in \mathbb{R}^{P}$ of all zeros except at the index corresponding to the control action in $\mathcal{A}$ that gives the lowest expected cost, where it is one.
See \cite{thorpe2021stochastic} for more details.
\section{Numerical Results}
\label{section: numerical results}

In all experiments, we use a Gaussian kernel function \(\kernelFunction(\state, \state') = \exp(- \lVert \state - \state' \rVert^{2} / 2 \gaussianBandwidth^{2})\), \(\gaussianBandwidth > 0\), and the hyperparameters \(\gaussianBandwidth\) and \(\regularizationParameter\) are chosen via cross-validation. 
See \cite{scholkopf2002learning, song2009hilbert} and \cite{li2022optimal} for a detailed discussion of parameter selection. 
Code to reproduce all experiments is available at \url{https://github.com/ajthor/socks}. 


\subsection{Spring-Mass-Damper System}
\label{sec:experiments_spring_mass}

\begin{figure*}[t]
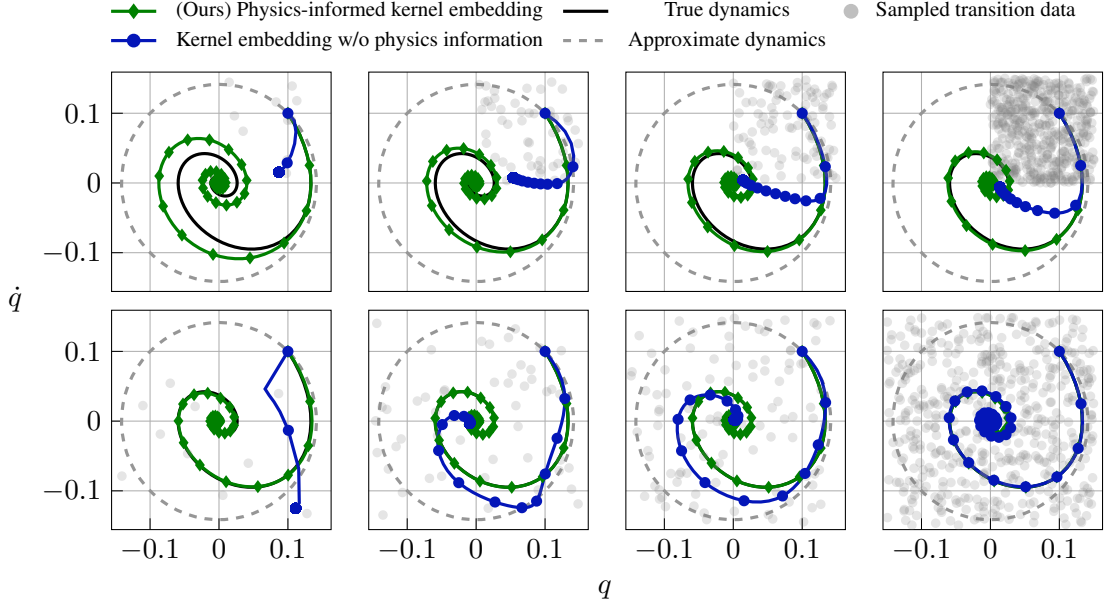

    \centering
    \vspace*{-0.5cm}
\begin{tikzpicture}

\definecolor{color0}{rgb}{1,0,1}
\definecolor{darkgray176}{RGB}{176,176,176}

\begin{customlegend}[
    legend columns=3, 
    legend style={
        align=left, 
        column sep=1ex, 
        font=\scriptsize, 
        draw=none,
        at={(130.0mm, 40.0mm)},
    }, 
    legend entries={(Ours) Physics-informed kernel embedding, True dynamics, Sampled transition data, Kernel embedding w/o physics information, Approximate dynamics}
]
\addlegendimage{very thick, learnedWithSideInfo, mark=diamond*, mark size=2, mark options={solid}}
\addlegendimage{very thick, trueDynamics}
\addlegendimage{semithick, gray, opacity=0.5, mark=*, mark size=2.5, mark options={solid}, only marks}
\addlegendimage{very thick, learnedWithoutSideInfo, mark=*, mark size=2, mark options={solid}}
\addlegendimage{very thick, approxDynamics, dashed}
\end{customlegend}

\begin{groupplot}[group style={group name = plots, group size=4 by 2, vertical sep=0.25cm, horizontal sep=0.5cm}]

\nextgroupplot[
tick align=outside,
tick pos=left,
x grid style={darkgray176},
xmajorgrids,
xmin=-0.155847587436438, xmax=0.162662648409605,
xtick style={color=black},
y grid style={darkgray176},
ylabel style={rotate=-90},
ylabel style={xshift=0.4cm},
ymajorgrids,
ymin=-0.155585780739784, ymax=0.159016266465187,
ytick style={color=black},
height=4.5cm,
width=4.5cm,
xmajorticks=false,
tick align=inside,
]
\input{figures/smd_phase_plots/phase_plot_10_quarter_space_data}

\nextgroupplot[
tick align=outside,
tick pos=left,
x grid style={darkgray176},
xmajorgrids,
xmin=-0.155847587436438, xmax=0.162662648409605,
xtick style={color=black},
y grid style={darkgray176},
ymajorgrids,
ymin=-0.155585780739784, ymax=0.159016266465187,
ytick style={color=black},
height=4.5cm,
width=4.5cm,
ticks=none,
tick align=inside,
]
\input{figures/smd_phase_plots/phase_plot_50_quarter_space_data}

\nextgroupplot[
tick align=outside,
tick pos=left,
x grid style={darkgray176},
xmajorgrids,
xmin=-0.155847587436438, xmax=0.162662648409605,
xtick style={color=black},
y grid style={darkgray176},
ymajorgrids,
ymin=-0.155585780739784, ymax=0.159016266465187,
ytick style={color=black},
height=4.5cm,
width=4.5cm,
ticks=none,
tick align=inside,
]
\input{figures/smd_phase_plots/phase_plot_100_quarter_space_data}

\nextgroupplot[
tick align=outside,
tick pos=left,
x grid style={darkgray176},
xmajorgrids,
xmin=-0.155847587436438, xmax=0.162662648409605,
xtick style={color=black},
y grid style={darkgray176},
ymajorgrids,
ymin=-0.155585780739784, ymax=0.159016266465187,
ytick style={color=black},
height=4.5cm,
width=4.5cm,
ticks=none,
tick align=inside,
]
\input{figures/smd_phase_plots/phase_plot_500_quarter_space_data}

\nextgroupplot[
tick align=outside,
tick pos=left,
x grid style={darkgray176},
xmajorgrids,
xmin=-0.155847587436438, xmax=0.162662648409605,
xtick style={color=black},
y grid style={darkgray176},
ymajorgrids,
ymin=-0.155585780739784, ymax=0.159016266465187,
ytick style={color=black},
ylabel=\(\canonicalVelocity\),
ylabel style={rotate=-90, at=(ticklabel cs:1.05), xshift=-0.4cm},
ylabel style={xshift=0.4cm},
height=4.5cm,
width=4.5cm,
tick align=inside,
]
\input{figures/smd_phase_plots/phase_plot_10_full_space_data}

\nextgroupplot[
tick align=outside,
tick pos=left,
x grid style={darkgray176},
xmajorgrids,
xmin=-0.155847587436438, xmax=0.162662648409605,
xtick style={color=black},
y grid style={darkgray176},
ymajorgrids,
ymin=-0.155585780739784, ymax=0.159016266465187,
ytick style={color=black},
xlabel=\(\canonicalCoord\),
xlabel style={at=(ticklabel cs:1.075)},
height=4.5cm,
width=4.5cm,
tick align=inside,
ymajorticks=false,
]
\input{figures/smd_phase_plots/phase_plot_50_full_space_data}

\nextgroupplot[
tick align=outside,
tick pos=left,
x grid style={darkgray176},
xmajorgrids,
xmin=-0.155847587436438, xmax=0.162662648409605,
xtick style={color=black},
y grid style={darkgray176},
ymajorgrids,
ymin=-0.155585780739784, ymax=0.159016266465187,
ytick style={color=black},
height=4.5cm,
width=4.5cm,
tick align=inside,
ymajorticks=false,
]
\input{figures/smd_phase_plots/phase_plot_100_full_space_data}

\nextgroupplot[
tick align=outside,
tick pos=left,
x grid style={darkgray176},
xmajorgrids,
xmin=-0.155847587436438, xmax=0.162662648409605,
xtick style={color=black},
y grid style={darkgray176},
ymajorgrids,
ymin=-0.155585780739784, ymax=0.159016266465187,
ytick style={color=black},
height=4.5cm,
width=4.5cm,
tick align=inside,
ymajorticks=false,
]
\input{figures/smd_phase_plots/phase_plot_500_full_space_data}

\end{groupplot}

\end{tikzpicture}
    \vspace*{-0.5cm}
    \caption{
        Phase-space trajectories predicted by the physics-informed kernel embedding (green) using imperfect dynamics (gray dashed), and by a purely data-driven embedding (blue). Top row: our approach accurately predicts the system behavior, despite having imperfect system knowledge and data from a limited region of the state space. Bottom row: our approach demonstrates better performance with smaller sample sizes. 
    }
    \vspace*{-0.5cm}
    \label{fig:smd_phase_plots}
\end{figure*}

For the purpose of analysis, we first consider the prediction problem in \eqref{eqn: prediction problem} with an (uncontrolled) spring-mass-damper system.
The equations of motion are given by 
$m \ddot{q} = - b \dot{q} - k q$.
We presume that we have access to \emph{imperfect} system dynamics $\tilde{f}(x) = -(k/m) q$, corresponding to an undamped spring-mass system. 
We generate a synthetic dataset $\mathcal{S} = \lbrace (x_{i}, y_{i}) \rbrace_{i=1}^{M}$ with varying sample sizes $M = 10, 50, 100, 500$, where the states $x_{i}$ are taken randomly from a bounded region of $\mathcal{X}$, and $y_{i} = f(x_{i})$ are corresponding next states at the subsequent timestep. 
We consider two cases for the sample: 1) the states $x_{i}$ are taken within the region $[0, 0.15] \times [0, 0.15]$, meaning we only have information within a limited operating regime, and 2) the states are taken within the region $[-0.15, 0.15] \times [-0.15, 0.15]$, which fully encompasses the operating region.


Using the sample $\mathcal{S}$, we then compute the physics-informed kernel embedding $\hat{m}_{0}$ using \eqref{eqn: biased regularized least squares problem} with $\sigma = 0.2$, and use $\hat{m}_{0}$ to predict the system evolution via \eqref{eqn: prediction problem} over $N = 100$ time steps from a fixed initial condition $x_{0} = [0.1, 0.1]^{\top}$.
To provide a baseline for comparison, we also use the purely data-driven embedding $\hat{m}$, computed using $\mathcal{S}$ via \eqref{eqn: regularized least squares problem} \cite[see][]{thorpe2021stochastic} to compute \eqref{eqn: prediction problem}.

The top row of Figure \ref{fig:smd_phase_plots} shows the performance of our approach for sample sizes $M = 10$, $50$, $100$, $500$ when data is collected from a limited region of the state space. 
Our approach demonstrates good empirical performance, and accurately predicts the evolution of the system despite having imperfect knowledge based on the undamped system under a wide range of conditions. 
As expected, the purely data-driven prediction does not accurately predict the system evolution outside the data region, even as the amount of data increases (top right plot). 
When data is collected over the entire region of interest, the quality of the purely data-driven estimate improves as the amount of data increases (bottom row of Figure \ref{fig:smd_phase_plots}). 
Note that our proposed approach has sound performance even while using only a small fraction of the data.
We note the following important trends. 

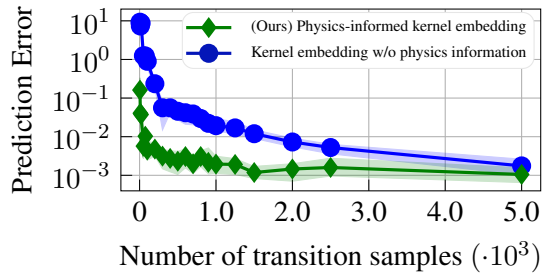
\begin{wrapfigure}{R}{0.5\textwidth}
    \centering
    \vspace*{-0.5cm}
\begin{tikzpicture}

\definecolor{darkgray176}{RGB}{176,176,176}
\definecolor{green}{RGB}{0,128,0}


\begin{axis}[
log basis y={10},
tick pos=left,
x grid style={darkgray176},
xlabel={Number of transition samples \((\cdot 10^{3})\)},
xmajorgrids,
xmin=-244.75, xmax=5249.75,
xtick style={color=black},
y grid style={darkgray176},
ymajorgrids,
ymin=0.000395890348402237, ymax=20.1672716452605,
ymode=log,
ytick style={color=black},
ytick={1e-05,0.0001,0.001,0.01,0.1,1,10,100,1000},
xtick={0, 1000, 2000, 3000, 4000, 5000},
xticklabels={0, 1.0, 2.0, 3.0, 4.0, 5.0},
title={},
ylabel={Prediction Error},
tick align=inside,
width=0.47*\textwidth,
height=4.0cm,
yticklabels={
  \(\displaystyle {10^{-5}}\),
  \(\displaystyle {10^{-4}}\),
  \(\displaystyle {10^{-3}}\),
  \(\displaystyle {10^{-2}}\),
  \(\displaystyle {10^{-1}}\),
  \(\displaystyle {10^{0}}\),
  \(\displaystyle {10^{1}}\),
  \(\displaystyle {10^{2}}\),
  \(\displaystyle {10^{3}}\)
}
]
\path [draw=blue, fill=blue, opacity=0.2]
(axis cs:5,12.3222302398093)
--(axis cs:5,6.78224948033399)
--(axis cs:10,5.89357859795471)
--(axis cs:20,6.8182997648858)
--(axis cs:50,0.894419479781164)
--(axis cs:75,0.762953791226013)
--(axis cs:100,0.602983344848617)
--(axis cs:200,0.182902450509996)
--(axis cs:300,0.0149416051880354)
--(axis cs:400,0.0417156344520063)
--(axis cs:500,0.0269613601072673)
--(axis cs:600,0.0209314235368356)
--(axis cs:700,0.0203387698399876)
--(axis cs:800,0.0166707086573566)
--(axis cs:900,0.0147656133487432)
--(axis cs:1000,0.0135893052361417)
--(axis cs:1250,0.0120723929521156)
--(axis cs:1500,0.00900157803238131)
--(axis cs:2000,0.00544862422566718)
--(axis cs:2500,0.00360110770501223)
--(axis cs:5000,0.00102015037709035)
--(axis cs:5000,0.00261650233063629)
--(axis cs:5000,0.00261650233063629)
--(axis cs:2500,0.00634236965302763)
--(axis cs:2000,0.00917567417252228)
--(axis cs:1500,0.0150829243201491)
--(axis cs:1250,0.024391355247763)
--(axis cs:1000,0.0291987394415522)
--(axis cs:900,0.031800240638725)
--(axis cs:800,0.0430077868430191)
--(axis cs:700,0.0533164440653225)
--(axis cs:600,0.0599015971638483)
--(axis cs:500,0.0685260305732231)
--(axis cs:400,0.0687455351446878)
--(axis cs:300,0.0822874910885258)
--(axis cs:200,0.29111067835569)
--(axis cs:100,1.11687103762178)
--(axis cs:75,1.63378131220179)
--(axis cs:50,1.73914324872963)
--(axis cs:20,9.89442017826479)
--(axis cs:10,8.48801357296558)
--(axis cs:5,12.3222302398093)
--cycle;

\path [draw=green, fill=green, opacity=0.2]
(axis cs:5,0.335940673981542)
--(axis cs:5,0.0965995571419611)
--(axis cs:10,0.018972433704367)
--(axis cs:20,0.0224030166058831)
--(axis cs:50,0.00267071865780281)
--(axis cs:75,0.00620288172003551)
--(axis cs:100,0.00247105764115584)
--(axis cs:200,0.00313253701719893)
--(axis cs:300,0.00146993378194456)
--(axis cs:400,0.0012318544184127)
--(axis cs:500,0.00105399956721517)
--(axis cs:600,0.00155283715842148)
--(axis cs:700,0.0011102397305128)
--(axis cs:800,0.00166865149024402)
--(axis cs:900,0.00108759458264064)
--(axis cs:1000,0.0011279049037076)
--(axis cs:1250,0.00105258498445505)
--(axis cs:1500,0.000824378780945961)
--(axis cs:2000,0.000704110177045369)
--(axis cs:2500,0.000957551146013936)
--(axis cs:5000,0.000647936943441525)
--(axis cs:5000,0.00156546567609673)
--(axis cs:5000,0.00156546567609673)
--(axis cs:2500,0.00276750761663653)
--(axis cs:2000,0.00199166095646712)
--(axis cs:1500,0.00169946262891485)
--(axis cs:1250,0.00270503544130059)
--(axis cs:1000,0.00313124971102351)
--(axis cs:900,0.00510445690467411)
--(axis cs:800,0.00428522385320571)
--(axis cs:700,0.00307608011272583)
--(axis cs:600,0.00483785161341719)
--(axis cs:500,0.00418954125102802)
--(axis cs:400,0.00431866642465888)
--(axis cs:300,0.00468335502349247)
--(axis cs:200,0.00684368215330715)
--(axis cs:100,0.00762181647981207)
--(axis cs:75,0.0140082738525924)
--(axis cs:50,0.0114727236853407)
--(axis cs:20,0.0498722111983917)
--(axis cs:10,0.0690606528037179)
--(axis cs:5,0.335940673981542)
--cycle;

\addplot [very thick, blue, mark=*, mark size=3, mark options={solid}]
table {%
5 9.21716214891041
10 7.3466051830649
20 8.70153437094929
50 1.24606273227491
75 1.23249437404531
100 0.895893729807677
200 0.235049973484926
300 0.0560164833286644
400 0.0558780568627893
500 0.0455226580428051
600 0.0417082228358015
700 0.0387466105916532
800 0.0295748522044077
900 0.0221927866215786
1000 0.0192842843299092
1250 0.0168737761773853
1500 0.0119339608871746
2000 0.00728888833207918
2500 0.00524147615694661
5000 0.00175161281193858
};
\addplot [very thick, green, mark=diamond*, mark size=3, mark options={solid}]
table {%
5 0.161198457965614
10 0.0405061811064313
20 0.0374247014641829
50 0.00569829978287408
75 0.010296390995401
100 0.00450854214920913
200 0.00477845991476066
300 0.0031368470133673
400 0.00273831627223761
500 0.00231887900802736
600 0.00305886796474358
700 0.00198993433572228
800 0.00306069383408771
900 0.00216299395288025
1000 0.00194988403045735
1250 0.0019120691208243
1500 0.00118025156340405
2000 0.00145759160663346
2500 0.00159407897624292
5000 0.00105042406009554
};
\end{axis}

\begin{customlegend}[
    legend columns=1, 
    legend style={
        align=center, 
        column sep=1ex, 
        font=\tiny, 
        draw=black!10!white,
        fill=white,
        inner xsep=0.3mm,
        inner ysep=0.3mm,
        rounded corners=1.5mm,
        at={(55.0mm, 23.8mm)},
    }, 
    legend entries={(Ours) Physics-informed kernel embedding, Kernel embedding w/o physics information}
]
\addlegendimage{very thick, learnedWithSideInfo, mark=diamond*, mark size=3, mark options={solid}}
\addlegendimage{very thick, learnedWithoutSideInfo, mark=*, mark size=3, mark options={solid}}
\end{customlegend}

\end{tikzpicture}

    \vspace*{-0.2cm}
    \caption{
        Physics-informed embeddings demonstrate lower empirical error than purely data-driven embeddings.
    }
    \vspace*{-1cm}
    
    \label{fig:smd_err_vs_num_data}
\end{wrapfigure}

\paragraph{Approximate physics knowledge improves out-of-distribution prediction accuracy.}
As shown in the top row of Figure \ref{fig:smd_phase_plots},
in contrast to the purely data-driven embedding, the physics-informed kernel embeddings generalize beyond the training dataset.

\paragraph{Approximate physics knowledge improves sample efficiency.}
As seen in the bottom row of Figure \ref{fig:smd_phase_plots}, when the observed transition data encompasses the entire region of interest, our approach is able to accurately predict the dynamics using only \(10\) data points.

\paragraph{Approximate physics knowledge reduces the prediction error.}
Figure \ref{fig:smd_err_vs_num_data} compares the empirical prediction error of the physics-informed kernel embedding $\hat{m}_{0}$ against the purely data-driven embedding $\hat{m}$. 
We randomly sample \(100\) initial states 
$x_{0}$ uniformly in the region $[-0.1, 0.1] \times [-0.1, 0.1]$,
and use the learned embeddings to predict the evolution of the state \(100\) time steps into the future.
Figure \ref{fig:smd_err_vs_num_data} shows the median
cumulative prediction error along these predicted trajectories (measured as the Euclidean distance between the true state vector and the predicted state vector).
We observe that for small datasets (particularly for \(\numSamples\) smaller than \(200\)) the physics-informed kernel embedding enjoys prediction error values that are two orders of magnitude smaller than those of the purely data-driven kernel embedding, 
and that the baseline method requires at least $5{,}000$ data points to achieve comparable levels of accuracy.


\subsection{F-16 Aircraft}
\label{section: f16 example}

We consider a ground collision avoidance scenario for an F-16 aircraft at initial altitude, as described in \cite{pmlr-v168-djeumou22b, heidlauf2018verification}. 
The underlying nonlinear dynamics, containing $13$ states and $4$ control inputs, capture the ($6$-DOF)
motion via evolution of
velocity $v_t$, angle of attack $\alpha$, sideslip $\beta$, altitude $h$, attitude angles: roll $\phi$, pitch $\theta$, yaw $\psi$, and their corresponding rates $p$, $q$, $r$, engine $power$ and two more states $p_n, p_e$ for translation along north and east, as in \cite{stevens2015aircraft}. 
The plant is built on linearly interpolated lookup tables that incorporate wind tunnel data describing the engine model, and other dynamic coefficients.
We inject zero-mean Gaussian noise with a standard deviation of $1\%$ of the magnitude of each state, such that the noise scales with the state magnitude.
We consider the case where the true dynamics are unknown, but presume that we have access to approximate dynamics with incorrect model parameters, including a gravitational constant of $g = 7.0$, and the interpolated lookup tables for the elevator control are half of their original values. 
These changes significantly alter the response of the aircraft to pulling up and avoiding collision with the ground. 

\begin{figure*}[t]
    \centering
    \vspace*{-0.5cm}
    \includegraphics[width=\linewidth]{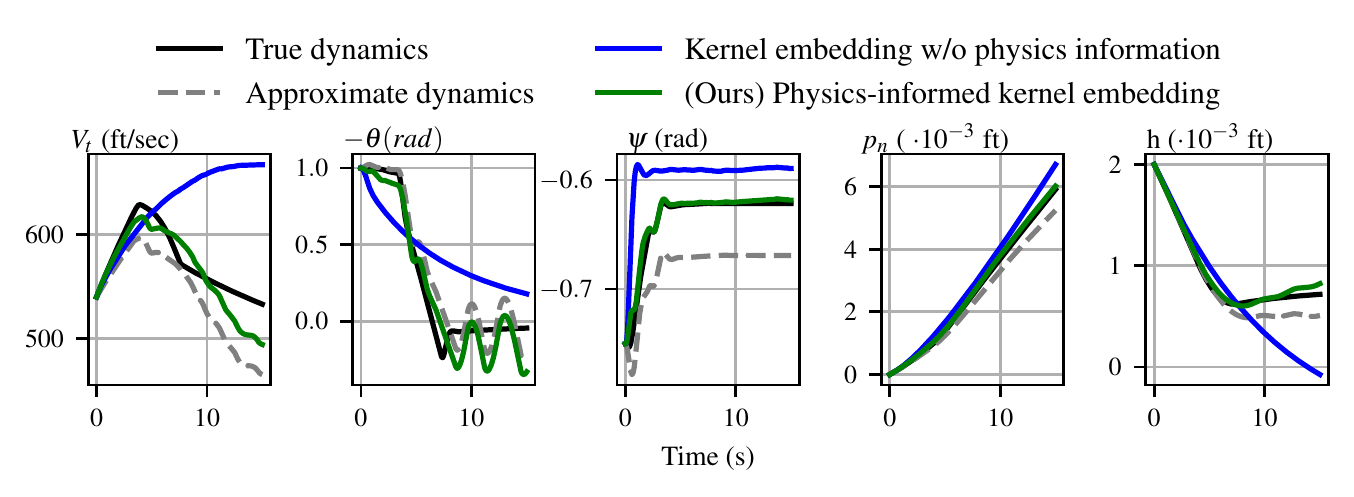}
    \vspace*{-1cm}
    \caption{
    Our approach, physics-informed kernel embeddings (green), accurately predicts the dynamics of an F-16 aircraft.
    The purely data-driven approach (blue) is inaccurate due to the system's high-dimensional and nonlinear dynamics.
    }
    \vspace*{-0.5cm}
    \label{fig:f16_expe}
\end{figure*}

We collect a sample $\mathcal{S} = \lbrace (x_{0,i}, \xi_{i}) \rbrace_{i=1}^{M}$, consisting of $M = 500$ initial conditions $x_{0, i}$ taken uniformly such that $[v_t, \alpha, \beta, \phi, \theta, p, q, r, h, power]^{\top} \in [490,590] \times  [-0.01, 0.09] \times [-0.05,0.05] \times [0.55,0.95] \times [-1.2,-0.8] \times [-0.2,0.2] \times [-0.2,0.2] \times [-0.2,0.2] \times [3800, 4200] \times [8.7, 9.3]$, and the resulting trajectories from those initial conditions $\xi_{i} \sim T(\cdot \mid x_{0, i})$ using the true, nominal dynamics, where $T : \mathscr{B}_{\mathcal{X}^{N}} \times \mathcal{X} \to [0, 1]$ is a stochastic kernel that represents the LQR-controlled, closed-loop system dynamics over $N = 1500$ time steps. 
Using trajectory data modifies the probability model to be a stochastic kernel over state trajectories, but does not significantly alter the kernel estimate.
Modifications of our approach to accommodate trajectory data is described in \cite{thorpe2022data}.

Figure \ref{fig:f16_expe} shows the solution to \eqref{eqn: prediction problem} for state prediction. 
We see significant improvement in prediction accuracy over the purely data-driven approach, in particular
the altitude $h$ and yaw angle $\Psi$. 
As expected, the purely data-driven method fails to capture the F-16 system behavior with the limited data due to the highly nonlinear and high-dimensional dynamics.
Interestingly, the prediction of the pitch angle $\theta$ using our approach shows oscillations due to the approximate dynamics. This raises question of whether the addition of data can overcome unstable model effects in the approximate dynamics. 
However, we leave this for future work. 


\subsection{Control of a Nonholonomic Vehicle System}
\label{section: nonholonomic example}

\begin{figure*}
    \centering
    \vspace*{-0.5cm}
    \begin{tikzpicture}
\begin{customlegend}[
    legend columns=4, 
    legend style={
        align=left, 
        column sep=1ex, 
        font=\scriptsize, 
        draw=none,
        at={(150.0mm, 38.0mm)},
    }, 
    legend entries={
        Sampled Transition Data,
        Target Trajectory,
        Data-Driven Trajectory, 
        (Ours) Physics-Informed Trajectory, 
       }
]
\addlegendimage{semithick, gray, opacity=0.3, mark=*, mark size=2.0, mark options={solid}, only marks}
\addlegendimage{very thick, mark=*, mark options={solid}, mark size=1.0, trueDynamics}
\addlegendimage{very thick, learnedWithoutSideInfo}
\addlegendimage{very thick, learnedWithSideInfo}
\end{customlegend}

\end{tikzpicture}
    \vspace*{-0.7cm}
    \includegraphics{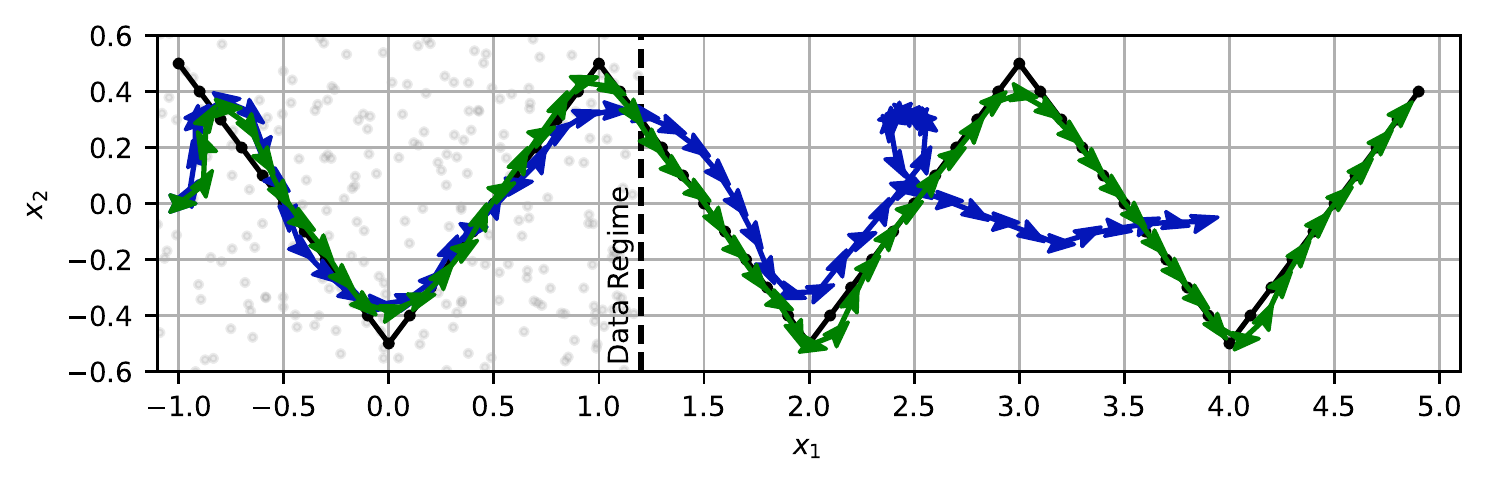}
    \vspace*{-0.0cm}
    \caption{Comparison of our proposed method against \cite{thorpe2021stochastic}. 
    The solution via physics-informed kernel embeddings (green) closely follows the target trajectory (black), even outside the data regime, while the performance of the purely data-driven solution (blue) degrades outside the region for which we have data.}
    \vspace*{-0.5cm}
    \label{fig: nonholonomic}
\end{figure*}

We solve \eqref{eqn: stochastic optimal control problem} for a target tracking control problem with a nonholonomic vehicle, as in \cite{thorpe2021stochastic}. The dynamics are given by
$\dot{x}_{1} = u_{1} \sin(x_{3})$, $\dot{x}_{2} = u_{1} \cos(x_{3})$, $\dot{x}_{3} = u_{2}$,
where $x = [x_{1}, x_{2}, x_{3}]^{\top} \in \mathbb{R}^{3}$ is the state and $u = [u_{1}, u_{2}]^{\top} \in \mathbb{R}^{2}$ is the control input, which we constrain to be within the bounds $[0.2, 1.5] \times [-10.1, 10.1]$.
We discretize the system in time and apply an affine disturbance with an exponential distribution $w_{t} \sim \mathrm{Exp}(0.1)$, with PDF $f(x; \alpha) = \alpha \exp(-\alpha x)$ if $x \geq 0$ and $f(x; \alpha) = 0$ if $x < 0$. We presume that the \emph{deterministic} discrete-time dynamics are given as approximate dynamical system knowledge, but that the \emph{stochastic} dynamics are unknown (i.e.\ we do not have prior knowledge of the disturbance).

We seek to solve \eqref{eqn: stochastic optimal control problem}, where we minimize the squared Euclidean distance to a moving target over a time horizon of $N = 60$.
We define a trajectory of target waypoints $z_{0}, z_{1}, \ldots, z_{N}$ (shown in black in Figure \ref{fig: nonholonomic}).
We consider the case where the future target position is unknown. Thus, we solve the following (unconstrained) optimization problem at each time step: $\min_{\pi} \mathbb{E}[\lVert x_{t+1} - z_{t} \rVert^{2}]$ as in \eqref{eqn: stochastic optimal control problem}. 
See \cite{thorpe2021stochastic} for more details.
We collect a sample $\mathcal{S} = \lbrace (x_{i}, u_{i}, y_{i}) \rbrace_{i=1}^{M}$ of size $M = 500$, where the states $x_{i}$ are taken uniformly in the region shown in Figure \ref{fig: nonholonomic}.
To compute the control algorithm in \eqref{eqn: approximate stochastic optimal control problem}, we generate a sample $\smash{\mathcal{A} = \lbrace \tilde{u}_{j} \rbrace_{j=1}^{P}}$ of $P = 210$ control actions taken uniformly in the region $[0.2, 1.2] \times [-10.1, 10.1]$.
We then presume that the true dynamics are unknown for the purpose of computing the control inputs.
We then computed the physics informed kernel embedding $\hat{m}_{0}$ with $\sigma = 0.75$.
Using $\hat{m}_{0}$, we simulate the system from an initial condition $x_{0} = [-1, 0, \pi/2]^{\top}$ and solve \eqref{eqn: approximate stochastic optimal control problem} at each time step to compute the stochastic policy. The total computation time was approximately $0.272$ seconds, and the results are shown in Figure \ref{fig: nonholonomic}. 
Using the same sample size, the baseline method from \cite{thorpe2021stochastic} fails to generate a meaningful trajectory (not shown). To generate a comparable trajectory, we used a much larger sample size, $M = 5000$, shown in blue in Figure \ref{fig: nonholonomic}, and the computation time was approximately $5.993$ seconds. 
This shows that our method demonstrates better empirical and computational performance, and requires less data due to the inclusion of prior dynamics knowledge.

\section{Conclusions \& Future Work}
\label{section: conclusion}

In this paper, we presented physics-informed kernel embeddings, a novel technique for incorporating prior system knowledge in data-driven representations of system dynamics using kernel distribution embeddings. 
Numerical experiments demonstrate the effectiveness of the proposed method on prediction tasks, including for systems with imperfect system knowledge on a spring-mass-damper system and highly nonlinear dynamics on an F-16 system, and on control tasks via a nonholonomic system target tracking problem. 
Results show that our approach generalizes well outside the data regime, is computationally efficient, and is robust to common sampling issues.

An important direction for future work in this area involves an exploration of how to incorporate other forms of prior knowledge, such as known system properties (e.g. symmetry, invariance) into the learning problem. Additionally, of practical interest is a characterization of the effect that poor or inaccurate approximate knowledge has on the learned representation.

\begin{acks}

This material is based upon work supported by the National Science Foundation under NSF Grants Number CNS-1836900 and NSF 1646522.  Any opinions, findings, and conclusions or recommendations expressed in this material are those of the authors and do not necessarily reflect the views of the National Science Foundation.
The NASA University Leadership initiative (Grant \#80NSSC20M0163) provided funds to assist the authors with their research, but this article solely reflects the opinions and conclusions of its authors and not any NASA entity.
This material is based upon work supported by the Air Force Office of Scientific Research under award number FA9550-19-1-0005.
Any opinions, findings, conclusions and or recommendations expressed in this material are those of the authors and do not necessarily reflect the views of the United States Air Force.
This material is based upon work supported by the Department of the Navy, Office of Naval Research under award number N00014-22-1-2254. 
Any opinions, findings, and conclusions or recommendations expressed in this material are those of the authors and do not necessarily reflect the views of the Office of Naval Research.

\end{acks}

\bibliography{bibliography}

\end{document}